\input harvmac
%
\rightline{RI-11-98}
\Title{
\rightline{hep-th/9811245}
}
{\vbox{\centerline{String Theory on $AdS_3 \times S^3 \times
S^3 \times S^1$}}}
\medskip
\centerline{\it Shmuel Elitzur, Ofer Feinerman, Amit Giveon, David Tsabar}
\vskip .3in
\centerline{Racah Institute of Physics}
\centerline{The Hebrew University}
\centerline{Jerusalem 91904, Israel}

\vskip .5in

\noindent
Spacetime properties of superstrings
on $AdS_3 \times S^3 \times S^3 \times S^1$ are studied.
The boundary theory is a two dimensional superconformal field
theory with a {\it large} $N=(4,4)$ supersymmetry.

\Date{11/98}

\def\journal#1&#2(#3){\unskip, \sl #1\ \bf #2 \rm(19#3) }
\def\andjournal#1&#2(#3){\sl #1~\bf #2 \rm (19#3) }

\def\frac#1#2{{#1\over#2}}

\def\half{\frac12}

\def\inbar{\,\vrule height1.5ex width.4pt depth0pt}
\def\IC{\relax\hbox{$\inbar\kern-.3em{\rm C}$}}
\def\IR{\relax{\rm I\kern-.18em R}}
\def\IP{\relax{\rm I\kern-.18em P}}

%
%

\def\cmp#1#2#3{Comm. Math. Phys. {\bf #1} (#2) #3}

\catcode`\@=11
\def\slash#1{\mathord{\mathpalette\c@ncel{#1}}}
\overfullrule=0pt

\def\GG{{\cal G}}

\def\LL{{\cal L}}
\def\MM{{\cal M}}
\def\NN{{\cal N}}

\def\TT{{\cal T}}

\def\underrel#1\over#2{\mathrel{\mathop{\kern\z@#1}\limits_{#2}}}

\catcode`\@=12


%

\def\exp{{\rm exp}}



\newsec{Introduction}

In this note we study the superstring propagation on the curved
spacetime manifold
\eqn\mm{\MM=AdS_3 \times S^3 \times S^3 \times S^1}
following
\ref\gks{A. Giveon, D. Kutasov and N. Seiberg, hep-th/9806194.}.
In string theory on $AdS_3\times \NN$ the spacetime theory
is a two dimensional conformal field
theory (CFT) on the boundary of $AdS_3$.
The left (right) moving affine $SL(2)$ symmetry on the worldsheet
(arising from the $AdS_3$ part of the background) is lifted to
a left (right) moving Virasoro algebra in the
spacetime boundary theory. Moreover, any affine
$\GG$ symmetry in the background $\NN$ CFT is lifted to an
affine $\GG$ algebra in the spacetime theory.

In section 2, we will show that type II string theory on $\MM$ \mm\
allows (the analogue) of a chiral GSO projection,
giving rise to a {\it large} $N=(4,4)$, $2$-$d$ SCFT
in spacetime. For simplicity, we usually
discuss the left moving sector of the
theory. The affine $SU(2)\times SU(2)\times U(1)$ (arising from
the $S^3 \times S^3 \times S^1$ background in \mm) is the
``R-symmetry'' of this large $N=4$ algebra. Hence, while string theory
on $AdS_3\times S^3\times M^4$, with $M^4=T^4$ or $K_3$, gives rise to a
{\it small} $N=4$ algebra in spacetime \gks\ (whose R-symmetry is $SU(2)$),
the example considered in this work provides a boundary theory
with the larger kind of $N=4$ supersymmetry in two dimensions.

\nref\ct{P.M. Cowdall and P.K. Townsend,
Phys. Lett. {\bf B429} (1998) 281, hep-th/9801165.}
\nref\bps{H.J. Boonstra, B. Peeters and K. Skenderis, hep-th/9803231.}
\nref\gmt{J.P. Gauntlett, R.C. Meyers and P.K. Townsend,
hep-th/9809065.}
Another reason to consider string propagation on
$\MM$ \mm\ is the following. The geometry
$AdS_3 \times S^3 \times S^3 \times R$ is obtained at the throat limit
of two differently oriented coincident sets
of fivebranes intersecting in one direction,
together with a set of infinitely stretched strings \refs{\ct,\bps,\gmt}.
Such background is the $S^1\rightarrow R$ limit
of $\MM$ in \mm.
Indeed, the study of Killing spinors on this space already
indicates the appearance of large $N=4$, $2$-$d$ supersymmetry \bps.

In section 3, we consider more aspects of the theory \mm,
and remark on some properties of the spacetime SCFT.

\newsec{Superstrings on $AdS_3\times S^3\times S^3\times S^1$}

\subsec{Worldsheet Properties}

As in \gks, we consider a fermionic string on
the Euclidean version of $AdS_3$, and study the
worldsheet theory in the free field representation
of $SL(2)$
\ref\wakim{M. Wakimoto, \cmp{104}{1986}{605}.}.
The worldsheet Lagrangian has the form
\eqn\wslag{\LL=\partial\phi\bar\partial\phi-
\sqrt{2\over k-2} R^{(2)}\phi+\beta\bar\partial
\gamma+\bar\beta\partial\bar\gamma-\beta\bar\beta\exp\left(
-\sqrt{2\over k-2}\,\phi\right)+\LL_{WZW}+\LL_{fer}}
The first part of $\LL$ is the bosonic part
of the $SL(2)$ WZW sigma model
(the target space of which is parametrized
by the coordinates $\phi$, $\gamma$, $\bar\gamma$ with metric
$ds^2=d\phi^2+e^{2\phi}d\gamma d\bar\gamma$,
and $\beta$ is an auxiliary
field inducing this metric in the sigma model upon integration).
The boundary of $AdS_3$ is at
$\phi\rightarrow\infty$, where the screening charge term vanishes,
and $(\gamma, \bar\gamma)$ are associated with the
coordinates parametrizing the two dimensional boundary.
The two point functions of the fields $\phi$, $\beta$, $\gamma$
are $\langle\phi(z)\phi(0)\rangle=-\log|z|^2$,
$\langle\beta(z)\gamma(0)\rangle=1/z$,
and all other two point functions vanish.
$\LL_{WZW}$ is the bosonic part of the $SU(2)\times SU(2)\times U(1)$
WZW model, and $\LL_{fer}$ includes all other terms
which, in particular, have worldsheet fermions
in them. There are three worldsheet fermions $\psi^A$ on the
$SL(2)$ manifold, three+three fermions $\chi^a$ and $\omega^a$
on $SU(2)\times SU(2)$, and a single fermion $\lambda$ on $U(1)$.
They are normalized such that
\eqn\normpsi{\eqalign{\langle\psi^A(z)\psi^B(w)\rangle
&=\frac{\eta^{AB}}{z-w}, \qquad A,B=1,2,3,
\qquad \eta^{AB}={\rm diag}(+,+,-) \cr
\langle\chi^a(z)\chi^b(w)\rangle
&=\frac{\delta^{ab}}{z-w}=\langle\omega^a(z)\omega^b(w)\rangle,
\qquad a,b=1,2,3\cr
\langle\lambda(z)\lambda(w)\rangle
&= \frac{1}{z-w}}}
The $SL(2)\times SU(2)\times SU(2)\times U(1)$ WZW
sigma model has affine bosonic currents
$j^A$, $k^a$, $m^a$, $\partial Y$ with levels $k+2$, $k'-2$, $k''-2$,
respectively. The worldsheet $N=1$ supercurrent $G$ of this system is
\ref\kazama{Y. Kazama and H. Suzuki, Nucl. Phys. {\bf B321} (1989) 232.}
\eqn\supgen{\eqalign{G(z)=&\sqrt{\frac{2}{k}}(\eta_{AB}\psi^A j^B
-\frac{i}{6}\epsilon_{ABC}\psi^A\psi^B\psi^C)
+\sqrt{\frac{2}{k'}}(\chi^a k_a -\frac {i}{6} \epsilon_
{abc}\chi^a \chi^b \chi^c)\cr
& + \sqrt{\frac{2}{k''}}(\omega^a
m_a-\frac{i}{6}\epsilon_{abc}\omega^a \omega^b \omega^c)
 +\lambda \partial Y}}
The total affine $SL(2)\times SU(2)\times SU(2)$ currents
$J^A$, $K^a$, $M^a$
are the upper components (with respect to
the $N=1$ worldsheet supersymmetry \supgen\ and up to normalization)
of the fermions
$\psi^A$, $\chi^a$, $\omega^a$, respectively:
\eqn\kaz{\eqalign{& J^A=j^A-\frac{i}{2}
                    \epsilon^A_{\,\, BC}\psi^B \psi^C \cr
      & K^a=k^a-\frac{i}{2} \epsilon^a_{\,\, bc}\chi^b \chi^c \cr
      & M^a=m^a-\frac{i}{2} \epsilon^a_{\,\, bc}\omega^b \omega^c \cr}}
They obey affine $SL(2)$, $SU(2)$ and $SU(2)$ algebras at levels
$k$, $k'$ and $k''$, respectively:
\eqn\slope{J^A(z)J^B(w)=\frac {k\eta^{AB}/2}{(z-w)^2}+
\frac {i\eta_{CD}\epsilon^{ABC}J^D}{(z-w)}+\cdots ,\qquad   A,B,C,D=1,2,3}
\eqn\suope{K^a(z)K^b(w)=\frac{k'\delta^{ab}/2}{(z-w)^2}+\frac{i
\epsilon^{ab}{  }_cK^c}{(z-w)}+\cdots, \qquad      a,b,c=1,2,3}
\eqn\sudope{M^a(z)M^b(w)=\frac{k''\delta^{ab}/2}{(z-w)^2}+\frac{i
\epsilon^{ab}{  }_cM^c}{(z-w)}+\cdots }
Finally, the $U(1)$ affine current $\partial Y$ is the upper component
of $\lambda$ and it satisfies:
\eqn\uone{\partial_z Y\partial_w Y=\frac{1}{(z-w)^2}+\cdots}
The central charge $c$ of the $N=1$ worldsheet theory on
$AdS_3 \times S^3 \times S^3 \times S^1$ is
\eqn\wscc{c=\frac{3(k+2)}{k}+\frac{3}{2}+\frac {3(k'-2)}
{k'}+\frac{3}{2}+\frac {3(k''-2)}{k''}+\frac{3}{2}
+1+\frac{1}{2}}
Criticality of the fermionic string $c=15$
together with eq. \wscc\ gives rise to the relation:
\eqn\nagad{\frac{1}{k}=\frac{1}{k'}+\frac{1}{k''}}

\subsec{Spacetime Properties}

In string theory, affine worldsheet symmetries are
lifted into global symmetries in spacetime.
As shown in \gks, a novelty of string theory on
$AdS_3$ is that such symmetries on the worldsheet are
lifted to {\it infinite} dimensional symmetries in the boundary,
$2$-$d$ spacetime theory.
In particular, the affine $SL(2)$ worldsheet symmetry
is lifted to a Virasoro algebra whose spacetime generators $L_n$
correspond to the zero modes of worldsheet holomorphic operators,
which can be chosen to be presented
(in the worldsheet BRST cohomology) by \gks:
\eqn\virgen{L_n=-\oint\,dz\left[(1-n^2)J^3\gamma^n +\frac{n(n-1)}{2}J^-
\gamma^{n+1} + \frac {n(n+1)}{2} J^+ \gamma^{n-1}\right]}
They satisfy the Virasoro algebra:
\eqn\virasoro{[L_n,L_m]=(n-m)L_{n+m}+\frac{c_{st}}{12}(n^3-n)\delta_{n+m,0}}
with a spacetime central charge
\eqn\cst{c_{st}=6kp}
where $p$ is interpreted \gks\ as the number of infinitely
stretched fundamental strings at the boundary of $AdS_3$.

The affine worldsheet $SU(2)\times SU(2) \times U(1)$
symmetry is lifted
to an affine $SU(2) \times SU(2) \times U(1)$ algebra in
spacetime. The modes $T_n^a$ generating the first affine
$SU(2)$ symmetry in spacetime correspond to the zero modes of the
worldsheet holomorphic operators:
\eqn\fsu{T^a_n=\sqrt{\frac{k'}{2}}\oint\,dz\{G_{-\frac{1}{2}}
,\chi^a \gamma^n(z)\}
=\oint\,dz \, \gamma^n
\left(K^a-n\sqrt\frac{k'}{k}\chi^a(\psi^3-
\half \psi^{-}\gamma-\half \psi^{+}\gamma^{-1})\right)}
where $G_{-\frac{1}{2}}=\oint\,dz\, G(z)$, and $G(z)$ is given in \supgen.
They satisfy the algebra
\eqn\fstsu{[T^a_n,T^b_m]=i\epsilon^{ab}{  }_cT^c_{n+m}
+\frac{k'_{st}}{2}n\delta^{ab}\delta_{n+m,0}}
\eqn\mix{[L_m,T^a_n]=-nT^a_{n+m}}
with a spacetime level \gks\
\eqn\kst{k'_{st}=k'p}
Similarly, the modes $R_n^a$ of the second $SU(2)$ algebra correspond to
\eqn\ssu{R^a_n=\sqrt{\frac{k''}{2}}\oint\,dz\{G_{-\frac{1}{2}}
,\omega^a \gamma^n(z)\}}
They satisfy the same algebra as \fstsu\ with a spacetime level:
\eqn\ksts{k''_{st}=k''p}
Finally, the affine $U(1)$ algebra has modes $\alpha_n$ corresponding
to
\eqn\uo{\alpha_n=\oint\,dz\{G_{-\frac{1}{2}},\lambda \gamma^n(z)\}
=\oint\, dz\, \gamma^n \left(\partial Y -
n\sqrt{\frac{2}{k}}\lambda(\psi^3-\half
\psi^{-}\gamma-\half \psi^{+}\gamma^{-1})\right)}
They satisfy the algebra
\eqn\alal{[\alpha_n,\alpha_m]=pn\delta_{n+m,0}}

Next we perform a chiral GSO projection,
thus making the spacetime theory supersymmetric.
To construct the spacetime supercharges
we introduce the 32 worldsheet spin fields $S^{\alpha}$,
$\alpha=1,...,32$, which satisfy
\ref\fms{D. Friedan, E. Martinec and S. Shenker, Nucl. Phys. {\bf B271}
(1986) 93.}:
\eqn\spsi{\eqalign{
&\psi^A(z)S_{\alpha}(w)
 ={1\over \sqrt{2}(z-w)^{1/2}}
  \Gamma^A_{\alpha\beta}S^{\beta}(w)+\cdots \cr
&\chi^a(z)S_{\alpha}(w)
 ={1\over \sqrt{2}(z-w)^{1/2}}
  \Gamma^{3+a}_{\alpha\beta}S^{\beta}(w)+\cdots \cr
&\omega^a(z)S_{\alpha}(w)
 ={1\over \sqrt{2}(z-w)^{1/2}}
  \Gamma^{6+a}_{\alpha\beta}S^{\beta}(w)+\cdots \cr
&\lambda(z)S_{\alpha}(w)
 ={1\over \sqrt{2}(z-w)^{1/2}}
  \Gamma^{10}_{\alpha\beta}S^{\beta}(w)+\cdots \cr
}}
where $\Gamma^i$, $i=1,...,10$, are 32 dimensional Dirac matrices
representing the Clifford algebra:
\eqn\clif{\{\Gamma^i,\Gamma^j\}=2g^{ij}I,
\qquad g^{ij}={\rm diag}(1^2,-1,1^7)}
A spacetime supersymmetry generator $Q$ has the
form \fms:
\eqn\superch{Q=\oint\,dz\, e^{-\frac{\varphi}{2}}S(z)}
where $\varphi$ is the scalar field
arising in the bosonized superghost system of the fermionic string,
and $S=\sum_{\alpha=1}^{32}u_{\alpha}S^{\alpha}$ is
a linear combination of spin fields.

To be physical,
$Q$ has to satisfy two conditions: it has to be BRST invariant and
it has to pass the GSO projection guaranteeing mutual locality amongst
spacetime fermions.
The BRST invariance reduces to the condition that
$S(z)$ have no $(z-w)^{-3/2}$ singularity
when contracted with $G(w)$ \supgen.
Using the OPEs of \spsi, this becomes
the following constraint on the
coefficients $u_{\alpha}$ defining $Q$:
\eqn\guz{\Gamma_{\alpha\beta}u^{\beta}=0}
where the 32 dimensional matrix $\Gamma$ is given by
\eqn\ggg{\Gamma=
         \sqrt{1\over k}(\Gamma^1\Gamma^2\Gamma^3)
        +\sqrt{1\over k'}(\Gamma^4\Gamma^5\Gamma^6)
        +\sqrt{1\over k''}(\Gamma^7\Gamma^8\Gamma^9)}
The matrix $\Gamma$ satisfies
\eqn\ggo{\Gamma^2=\left({1\over k}-{1\over k'}-{1\over k''}\right)I}
Equation \guz\ has a solution if and only if the criticality
condition \nagad\ is satisfied; in that case, one has $\Gamma^2=0$.
The hermitian conjugate of $\Gamma$,
\eqn\gher{\Gamma^{\dagger}=
         \sqrt{1\over k}(\Gamma^1\Gamma^2\Gamma^3)
        +\sqrt{1\over k'}(-\Gamma^4\Gamma^5\Gamma^6)
        +\sqrt{1\over k''}(-\Gamma^7\Gamma^8\Gamma^9)}
obeys also $(\Gamma^{\dagger})^2=0$, and we have
\eqn\ggdag{\{\Gamma,\Gamma^{\dagger}\}={2\over k}I}
The matrices $\sqrt{k/2}\,\Gamma$ and $\sqrt{k/2}\,\Gamma^{\dagger}$
are subjected to the algebra of fermionic annihilation and
creation operators. The 32 dimensional space splits into two 16
dimensional spaces corresponding to occupation number 0 or 1 for that
fermion. The space of solutions of \guz\ is then the 16 dimensional
occupation 0 subspace.

As in the flat space case, an appropriate GSO projection corresponds to the
requirement:
\eqn\geuu{\Gamma^{11}_{\alpha\beta}u^{\beta}=u_{\alpha},
\qquad \Gamma^{11}=\prod_{i=1}^{10}\Gamma^i}
Since $\Gamma^{11}$ anticommutes with $\Gamma$ of eq. \ggg,
it preserves its space of solutions reducing the number of
spacetime supercharges to 8. Under the $SO(9,1)$ rotating the
$\psi^i$, the spin fields transform with the spinor representation.
The condition \guz\ breaks this symmetry down to
$SL(2)\times SU(2)\times SU(2)$, under which the 8
supersymmetry generators transform in the
$(\bf{1/2},\bf{1/2},\bf{1/2})$ representation.

It is convenient to work with a bosonized form of the
spin fields $S^{\alpha}$. A particularly simple bosonized form for
the solutions of \guz\ and \geuu\ is obtained when defining the bosonic
fields according to the following choice of complex structure on $\MM$:
\eqn\bsnize{\eqalign{& \partial H_1=\psi^1 \psi^2 \cr
                     & \partial H_2=\chi^1 \chi^2 \cr
                     & \partial H_3= \omega ^1 \omega^2 \cr
   &i\partial H_4= \Big(\sqrt{\frac{k}{k'}}\chi^3+\sqrt{\frac{k}{k''}}
                        \omega^3\Big)\psi^3 \cr
   & \partial H_5= \Big(\sqrt{\frac{k}{k''}}\chi^3-\sqrt{\frac{k}{k'}}
                        \omega^3\Big)\lambda \cr}}
The five scalars $H_I$ are normalized such that
\eqn\hh{\langle H_I(z)H_J(w)\rangle=-\delta_{IJ}\log{(z-w)}}
The eight solutions of eqs. \guz\ and \geuu\ have then the form
\eqn\four{\eqalign{& S_{+++}=e^{\frac{i}{2}(H_1+H_2+H_3+H_4+H_5)} \cr
                   & S_{+--}=e^{\frac{i}{2}(H_1-H_2-H_3-H_4-H_5)} \cr
                   & S_{-++}=e^{\frac{i}{2}(-H_1+H_2+H_3-H_4+H_5)} \cr
                   & S_{---}=e^{\frac{i}{2}(-H_1-H_2-H_3+H_4-H_5)} \cr}}

\eqn\fourmore{\eqalign{
& S_{+-+}=\sqrt{\frac{k}{k'}}e^{\frac{i}{2}(H_1-H_2+H_3-H_4+H_5)}
        +\sqrt{\frac{k}{k''}}e^{\frac{i}{2}(H_1-H_2+H_3+H_4-H_5)} \cr
& S_{++-}= \sqrt{\frac{k}{k'}}e^{\frac{i}{2}(H_1+H_2-H_3+H_4-H_5)}
+\sqrt{\frac{k}{k''}}e^{\frac{i}{2}(H_1+H_2-H_3-H_4+H_5)}  \cr
&S_{--+}=\sqrt{\frac{k}{k'}}e^{\frac{i}{2}(-H_1-H_2+H_3+H_4+H_5)}
       +\sqrt{\frac{k}{k''}}e^{\frac{i}{2}(-H_1-H_2+H_3-H_4-H_5)} \cr
&S_{-+-}= \sqrt{\frac{k}{k'}}e^{\frac{i}{2}(-H_1+H_2-H_3-H_4-H_5)}
+\sqrt{\frac{k}{k''}}e^{\frac{i}{2}(-H_1+H_2-H_3+H_4+H_5)} \cr}}
The four solutions in \fourmore\ can be obtained,
for instance, by applying the
global generators of $SU(2)\times SU(2)$, $T_0^a$ and $R_0^a$ of eqs.
\fsu\ and \ssu, to the four simple solutions \four.

Notice that in the limit $k'$ or $k''$ tending to infinity,
both the complex structure \bsnize\ and the spin fields \four, \fourmore\
are identical to those used in \gks, corresponding to string theory on
$AdS_3\times S^3\times T^4$. Indeed, in the limit of large level an $SU(2)$
affine algebra approaches a $U(1)^3$ one.

Altogether, we have found eight left moving supercharges,
thus leading to a global $N=4$ supersymmetry in the NS sector
of the spacetime theory.
Indeed, the algebra generated by these eight physical supercharges,
together with $L_{0}$, $L_{\pm 1}$, $T^a_0$, $R^a_0$, is a two dimensional
global $N=4$ algebra with an $SU(2)\times SU(2)$ subalgebra;
explicitly:
\eqn\zero{\eqalign{ & [L_m,L_n]=(m-n)L_{m+n},\qquad m,n=0,\pm 1 \cr
                    & [T^a_0,T^b_0]=i\epsilon^{ab}{  }_c T^c_0, \qquad
                      [R^a_0,R^b_0]=i\epsilon^{ab}{  }_c R^c_0, \qquad
                      a,b=1,2,3\cr
                    & [T^a_0,R^b_0]=[L_n,T^a_0]=[L_n,R^a_0]=0  \cr
     & [L_n,G_r^{ij}]= \left(\frac{n}{2}-r\right) G_{n+r}^{ij} , \qquad
       i,j=\pm 1/2 , \qquad r,s=\pm 1/2    \cr
     & [T^a_0,G^{ij}_r]= \half (\sigma^a)^i_{\, k} G^{kj}_r, \qquad
       [R^a_0,G^{ij}_r]=\half  (\sigma^a)^j_{\, k} G^{ik}_r  \cr
     & \{ G^{ij}_r,G^{kl}_s \}=2\epsilon^{ik}\epsilon^{jl}L_{r+s}
       -2(r-s)\left(\frac{k}{k'}\epsilon^{jl}(\sigma_a)^{ik}T^a_0
       +\frac{k}{k''}\epsilon^{ik}(\sigma_a)^{jl}R^a_0\right) \cr }}
Here $G=i(2k)^{\frac{1}{4}}Q$
is normalized to satisfy
the last anticommutator, and the labels on $G_r^{ij}$
are identified with: $r\equiv \half \epsilon_1$,
$i\equiv\half \epsilon_2$ and $j\equiv\half \epsilon_3$.
In the limits $k' \rightarrow \infty$ or $k'' \rightarrow
\infty$, this algebra reduces to
the global, small $N=4$ algebra (and a $U(1)^3$ factor).

\nref\stv{A. Sevrin, W. Troost, A. Van Proeyen, Phys. Lett.
          {\bf B208} (1988) 447.}
\nref\sch{K. Schoutens, Nucl. Phys. {\bf B295} (1988) 634.}

Combining the global $N=4$ algebra \zero\ with the
full Virasoro algebra \virasoro\
and affine $SU(2)\times SU(2)\times U(1)$, \fstsu, \mix, \alal\
(in the worldsheet BRST cohomology and using the freedom
of picture changing \fms)
leads to the so called ``large'' $N=4$ superconformal algebra
\refs{\stv,\sch}.  This chiral algebra is generated by a
spin-2 stress tensor (whose modes are $L_n$, $n\in Z$),
four spin-3/2 fields (with modes $G_r^{ij}$, $r\in Z+1/2$),
seven spin-1 currents (with modes $T_n^a$, $R_n^a$, $\alpha_n$),
and four spin-1/2 fermions (with modes $\Gamma_r^{ij}$).
For instance, the modes $G_r^{ij}$ of the
four spacetime supercurrents can be obtained by
considering the products of the integrands in $L_n$ \virgen\
with the eight spin fields \four, \fourmore.
Similarly, the modes $\Gamma_r^{ij}$ of the four spacetime
fermions -- the $N=4$ superpartners of the $U(1)$
spacetime boson -- can be obtained by considering the products of
these eight physical spin fields with the $\alpha_n$ \uo.

The left movers on the worldsheet are lifted to left movers
in the spacetime boundary theory, while worldsheet right movers
are lifted to right movers in spacetime.
All in all, the spacetime theory is a two dimensional
SCFT with a large $N=(4,4)$ supersymmetry, with
a central charge $c_{st}$ \cst\ and levels $k'_{st}$, $k''_{st}$
\kst, \ksts. Using \nagad, we see that the spacetime
theory obeys the condition
\eqn\stcon{c_{st}={6k'_{st}k''_{st}\over k'_{st}+k''_{st}}}
as required in a unitary large $N=4$, $2$-$d$ theory \refs{\stv,\sch}.

\newsec{Remarks}

As shown in \gks, the properties of physical operators on
the worldsheet are naturally lifted into properties
of their corresponding physical states in the boundary
spacetime theory. For instance, the $SL(2)$ quantum number
$j$ and worldsheet spin $s$ are related to the scaling
dimension $h_{st}$ and spin $s$ in the spacetime $2$-$d$ theory.
Moreover, operators in a representation $R$ of the affine
$SU(2)\times SU(2)\times U(1)$ worldsheet symmetry are
lifted to states with similar properties in spacetime.
Instead of being general (for details see \gks) we will
discuss some particularly interesting examples.

Consider first the worldsheet vertex operators
\eqn\vertex{V(j)=e^{-\varphi-\bar\varphi}(\psi^3-\frac12\gamma\psi^{-}
-\frac12\gamma^{-1}\psi^{+})(\bar\psi^3-\frac12\bar\gamma\bar
\psi^{-}-\frac12\bar\gamma^{-1}\bar\psi^{+})V_{jm\bar m}
V'_{jm'\bar m'}V''_{jm''\bar m''}}
where $V_{jm\bar m}$, $V'_{jm'\bar m'}, V''_{jm''\bar m''}$
are vertex operators of the $SL(2)\times SU(2)\times SU(2)$
WZW model with isospins ($j,j^3=m,\bar j^3=\bar m$),
($j,k^3=m',\bar k^3=\bar m'$) and ($j,m^3=m'',\bar m^3=\bar m''$),
respectively.
The worldsheet scaling dimensions of $V(j)$ are
\eqn\scdi{h=\bar h={1\over 2}+{1\over 2}+
j(j+1)\left(-{1\over k}+{1\over k'}+{1\over k''}\right)=1}
(the last equality is obtained using the criticality condition \nagad),
and they are BRST invariant. Furthermore, the spacetime states
corresponding to $V(j)$ are primaries of the Virasoro algebra
with scaling dimensions $h_{st}=\bar h_{st}=j$, as well as primaries of
the spacetime affine $SU(2)\times SU(2)\times U(1)$ algebra.
Therefore, the physical states corresponding to $V(j)$
are chiral primaries of the $N=(4,4)$ superconformal algebra.
For $j=1/2$, the spacetime upper components of $V(1/2)$ include a
singlet of $SU(2)\times SU(2)\times U(1)$ with scaling dimensions
$h_{st}=\bar h_{st}=j+1/2=1$, which thus correspond to
(the $m, \bar m$ modes of) a single
marginal operator in the spacetime theory which preserves $N=(4,4)$
supersymmetry. On the worldsheet,
this marginal deformation is in the R-R sector and, therefore,
turns on a R-R background.

Together with the manifest marginal deformation changing the radius
of $S^1$, we thus see that the superstring on $\MM$ \mm\ has
a two dimensional moduli space.
Let us discuss the structure of this moduli space in more details.

The theory studied in section 2 is a WZW model on $\MM$;
it has a vanishing R-R background. In particular,
its $S^1\rightarrow R$ limit is obtained at the throat limit of
(appropriate smearing of strings in) the following system
\refs{\ct,\bps,\gmt}:
\item{(a)}
$k'$ coincident NS-fivebranes stretched, say, in the
$(x^0,x^1,x^2,x^3,x^4,x^5)$ directions.
\item{(b)}
$k''$ coincident NS-fivebranes stretched in the
$(x^0,x^1,x^6,x^7,x^8,x^9)$ directions.
\item{(c)}
$p$ fundamental strings infinitely stretched in $(x^0,x^1)$.

\noindent
In the type IIB string, this system is S-dual to a corresponding
configuration of D5-branes and D-strings.
The marginal deformation discussed above may very well be
the one which connects continuously the theories with vanishing
R-R backgrounds to those which have them turned on
(instead of the WZ term in the NS-NS sector).

It would be interesting to understand the structure of the $2$-$d$,
$N=(4,4)$ spacetime theory. Although this
is beyond the scope of this work,
let us speculate about the particular case $k'=k''=2k$.
In this case, the central charge \cst\ can be any positive
integer product of three:
$c_{st}=3n$, $n=k'p$. A large $N=4$ theory with $c=3$
is constructed out of a single scalar field and four fermions;
we denote such a theory by $\TT_3$.
A candidate for the spacetime CFT is thus a
(deformation of) an orbifold sigma model like
the symmetric product $(\TT_3)^{n}/S_{n}$.
Indeed, this is an $N=(4,4)$ SCFT with a single modulus
in the untwisted sector, corresponding
to the radius of the scalar in $\TT_3$, and another modulus
in the twisted sector which we found amongst the $S_3$ sector,
invariant twist fields~\foot{Note that for $n\leq 2$, unitarity implies that
\vertex\ are not in the physical spectrum (see \gks, and references
therein, for details), even for $j=1/2$.
Therefore, there is a perfect agreement with the fact that
$(\TT_3)^2/Z_2$ does not have a marginal deformation in the twisted
sector.}.

States in the $(\TT_3)^{n}/S_{n}$ SCFT appear in $N=(4,4)$ multiplets
and fall into $SU(2)\times SU(2)$ representations
\eqn\reps{\left({l+q\over 2}, {l-q\over 2}\right),
\qquad l,q\in {Z\over 2}, \quad 0\leq l-q\in Z}
Amongst the operators in a representation \reps, the one with the smallest
scaling dimension appears in the $Z_{2l+1}$ twisted sector.
The scaling dimension of such a twist field can be obtained by standard
considerations of the symmetric orbifold:
\eqn\horb{h_{orb}\left({l+q\over 2}, {l-q\over 2}\right)=
{l+\delta\over 2}, \qquad \delta={q^2\over 2l+1}}

In string theory on $AdS_3\times SO(4)\times U(1)$, worldsheet
operators corresponding to spacetime, primary states with
lowest scaling dimensions $h_{st}$, $\bar{h}_{st}$
in a given representation \reps\ are
\eqn\verjj{V(j',j'')=e^{-\varphi-\bar\varphi}(\psi^3-\frac12\gamma\psi^{-}
-\frac12\gamma^{-1}\psi^{+})(\bar\psi^3-\frac12\bar\gamma\bar
\psi^{-}-\frac12\bar\gamma^{-1}\bar\psi^{+})V_{jm\bar m}
V'_{j'm'\bar m'}V''_{j''m''\bar m''}}
with
\eqn\jpjp{(j',j'')=\left({l+q\over 2}, {l-q\over 2}\right)}
\eqn\jjj{j=-{1\over 2}+\sqrt{{1\over 4}+{j'(j'+1)+j''(j''+1)\over 2}}}
and
\eqn\hstj{h_{st}=\bar{h}_{st}=j}
We thus find that upon deforming the orbifold SCFT to the spacetime
SCFT corresponding to string theory on $AdS_3\times SO(4)\times U(1)$,
the dimensions of smallest $h$ operators in a representation \reps\
are changed as
\eqn\dh{(h_{orb}+1/2)^2-(h_{st}+1/2)^2={1\over 4}\delta(\delta+1)}
where $\delta$ is given in \horb.

For $\delta=0$, the dimensions of such operators are not changed.
Indeed, such operators are in the $(l/2,l/2)$ representation of
$SO(4)$ and their scaling dimension is $l/2$;
they are the chiral primaries \vertex\ discussed above.
For $\delta\neq 0$, the dimensions are changed when one deforms
away from the orbifold point by a relatively small amount \dh.

Finally, we should mention that the generalization of this work
to models with a large $N=(4,0)$ spacetime supersymmetry
can be done along the lines of \ref\llk{D. Kutasov, F. Larsen and R.G.
Leigh, to appear.}.

\bigskip
\noindent{\bf Acknowledgements:}
We thank E. Rabinovici, A. Schwimmer, N. Seiberg,
and especially D. Kutasov for very useful discussions.
This work is supported in part by the Israel
Academy of Sciences and Humanities -- Centers
of Excellence Program. The work of A.G.
is supported in part by BSF -- American-Israel Bi-National
Science Foundation. S.E. and A.G. thank the theory division
at CERN where part of this work was done,
and the Einstein Center at the Weizmann Institute for partial support.

\listrefs
\end